\newif\ifsubmit
\submittrue

\newif\ifarxiv
\arxivtrue

\newif\ifdraft
\draftfalse

\documentclass{llncs}
\newif\iffinal
\finaltrue

\usepackage[T1]{fontenc}
\usepackage[utf8]{inputenc}
\usepackage{lmodern}
\usepackage[scaled=.8]{beramono}
\usepackage{latexsym}
\usepackage{textcomp}

\usepackage{amssymb}
\usepackage[final]{graphicx}
\usepackage{caption}
\usepackage{listings}
\usepackage{booktabs}
\usepackage{mathtools}
\usepackage{tabularx}

\usepackage{tikz}
\ifsubmit
\usetikzlibrary{arrows,calc,positioning,shapes}
\else
\usetikzlibrary{arrows,calc,positioning,shapes}
\fi

\usepackage[ngerman,british]{babel}
\usepackage[inline]{enumitem}
\usepackage[babel]{csquotes}
\MakeAutoQuote{“}{”}
\MakeAutoQuote*{‘}{’}
\usepackage{soul}
\setul{.25ex}{}

\usepackage[obeyDraft,textsize=tiny]{todonotes}

\usepackage[backend=bibtex,hyperref=auto,
firstinits=true,style=numeric,
urldate=iso8601,isbn=false,doi=false]{biblatex}
\ifsubmit
\else
\fi
\DeclareNameAlias{author}{last-first}
\DeclareNameAlias{editor}{last-first}
\DeclareNameAlias{translator}{last-first}
\DeclareFieldFormat{labelnumberwidth}{#1.}
\DeclareFieldFormat{title}{#1}
\DeclareFieldFormat
  [article,inbook,incollection,inproceedings,patent,thesis,unpublished]
  {title}{#1}
\DeclareFieldFormat{journaltitle}{#1}
\newbibmacro*{volume+number+eid}{%
  \printfield{volume}%
  \iffieldundef{number}
    {}
    {(%
      \printfield{number}%
     )%
    }%
  \setunit{\addcomma\space}%
  \printfield{eid}}
\DeclareFieldFormat{url}{\url{#1}}

\usetikzlibrary{automata}

\begin{document}

\title{Semantic Publishing Challenge –\newline Assessing the Quality of Scientific Output by Information Extraction and Interlinking\ifarxiv\thanks{The final publication is available at \texttt{link.springer.com}}\fi}
\author{%
Angelo Di Iorio\inst{1}
\and Christoph Lange\inst{2}
\and Anastasia Dimou\inst{3}
\and Sahar Vahdati\inst{4}}
\institute{%
Università di Bologna, Italy
\email{diiorio@cs.unibo.it}
\and University of Bonn \& Fraunhofer IAIS, Germany
\email{math.semantic.web@gmail.com}
\and Ghent University – iMinds – Multimedia Lab, Belgium
\email{anastasia.dimou@ugent.be}
\and University of Bonn, Germany
\email{vahdati@uni-bonn.de}
}

\maketitle

\begin{abstract}
The Semantic Publishing Challenge series aims at investigating novel approaches for improving scholarly publishing using Linked Data technology.
In 2014 we had bootstrapped this effort with a focus on extracting information from non-semantic publications – computer science workshop proceedings volumes and their papers – to assess their quality.
The objective of this second edition was to improve information extraction but also to interlink the 2014 dataset with related ones in the LOD Cloud, thus paving the way for sophisticated end-user services.
\end{abstract}

\section{Introduction: Semantic Publishing Today}
\label{sec:schol-publ-semant}

The widely held assumption that “scholarly communication by means of semantically-enhanced media-rich digital publishing is likely to have a greater impact than [print or PDF]”~\cite{force11manifesto} is slowly coming true, pushed by regular events such as the workshop series on getting “Beyond the PDF”, semantic publishing and linked science\footnote{See \url{https://www.force11.org/meetings/beyond-pdf-2}, \url{http://sepublica.info}, and \url{http://linkedscience.org/category/workshop/}}.
Semantic technology is increasingly supporting researchers in disseminating, exploiting and evaluating their results using open formats.
Concrete technical solutions investigated by the semantic publishing community include:
\begin{itemize}
\item machine-comprehensible representations of scientific methods, models, experiments and research data,
\item links from papers to such data,
\item alternative publication channels (e.g.\ social networks and micro-publications),
\item alternative metrics for scientific quality and impact, e.g., taking into account the scientist's social network, user-generated micro-content such as discussion post, and recommendations.
\end{itemize}
Sharing scientific data and building new research on them will lead to data value chains increasingly covering the whole process of scientific research and communication.
The Semantic Publishing Challenges aim at supporting the buildup of such data value chains, initially by extracting information from non-semantic publications and interlinking this information with existing datasets.
Our prime use case is the computation of novel quality metrics based on such information.

Section~\ref{sec:chall-defin-publ} presents the definition of this year's Challenge, Section~\ref{sec:extraction-tasks} explains the evaluation procedure, Sections~\ref{sec:task1} to \ref{sec:task3} explain the definitions and outcomes of the three tasks in detail, and Section~\ref{sec:over-less-learn} discusses overall lessons
learnt.

\section{Definition of the Challenge}
\label{sec:chall-defin-publ}

In 2014, we had found it challenging to define a challenge about semantic publishing~\cite{LangeDiIorio:SemPub14}.
Existing datasets focused on basic bibliographical metadata or on research data specific to one scientific domain; we did not consider them suitable to enable advanced applications such as a comprehensive assessment of the quality of scientific output.
We had thus designed the first Challenge to produce, by information extraction and in an objectively measurable way, an initial data collection that would be useful for future challenges and that the community can experiment on.
As the two information extraction tasks had received few submissions, and as the community had asked for a more exciting task w.r.t.\ the future of scholarly publishing, we added an open task with a subjective evaluation.

In 2015, we left \textbf{Task~1} of 3 largely unchanged: answering queries related to the quality of workshops by computing metrics from data extracted from their proceedings, also considering information about persons and events.
The 2014 results had been encouraging, and we intended to give the 2014 participants an incentive to participate once more with improved versions of their tools.
As in 2014, \textbf{Task~2} focused on extracting contextual information from the full text of papers: citations, authors' affiliations, funding agencies, etc.
In contrast to 2014, we now used the same data source as for Task~1 (the CEUR-WS.org open access computer science workshop proceedings), to foster synergies between the two tasks and to encourage participants to compete in both tasks.
Based on the data obtained as a result of the 2014 Task~1, we defined the objective of \textbf{Task~3} to interlink the CEUR-WS.org linked data with other relevant linked datasets.

\section{Common Evaluation Procedures}
\label{sec:extraction-tasks}

The evaluation for all tasks followed a common procedure similar to the other Semantic Web Evaluation Challenges:\footnote{As no one participated in Task~3, our work on this task ended with step~\ref{it:queries}.}
\begin{enumerate}
\item\label{it:TD} For each task, we initially published a \emph{training dataset} (TD) on which the participants could test and train their extraction and interlinking tools.
\item\label{it:data-structure} For the information extraction tasks, we specified the basic structure of the RDF extracted from the TD source data, without prescribing a vocabulary.
\item\label{it:queries} We provided natural language queries and their expected results on TD.
\item\label{it:ED} A few days before the submission deadline, we published an \emph{evaluation dataset} (ED), a superset of TD, which was the input for the final evaluation.
\item\label{it:submission} We asked the participants to submit their linked data resulting from extraction or interlinking (under an open license to permit 
reuse%
), SPARQL implementations of 
the queries%
, as well as their extraction tools, as we reserved the right to inspect them.
\item\label{it:awards} We awarded prizes for the best-performing
(w.r.t.\ the F1 score computed from precision/recall)
and for the most innovative approach
(determined by the chairs\footnote{Anastasia Dimou, a co-author of one Task~1 submission~\cite{HeyEtAl:SemPub15}, did not vote in this task.}).
\item\label{it:transparency} Both before and after the submission we maintained transparency.
Prospective participants were invited to ask questions, e.g.\ about the expected query results, which we answered publicly.
After the evaluation, we made the scores and the gold standard (see below) available to the participants.
\end{enumerate}

\begin{figure}
  \centering
  \begin{minipage}{.57\textwidth}
    \begin{tikzpicture}
      \ifsubmit
      \ifarxiv
\begin{scope}[
  dataset/.style={draw,cylinder,shape border rotate=90,aspect=.25,minimum width=3em,minimum height=9ex,font={\scriptsize}},
  file/.style={draw,rectangle,minimum width=3em,minimum height=8ex,font={\scriptsize}},
  process/.style={->,
  },
  process-label/.style={align=center,font={\tiny}},
  align=center,
  node distance=.35cm]
  \node[dataset] (source) {Source\\ data};
  \node[dataset,right=of source] (lod) {Linked\\ data\\ (RDF)};
  \node[file,right=of lod] (result) {Query\\ result\\ (CSV)};
  \node[file,below=.5cm of result] (gold) {Gold\\ standard\\ (CSV)};
  \coordinate (mid) at ($(result.south)!.5!(gold.north)$);
  \node[file,right=1.5cm of mid] (report) {Report\\ (HTML)};
  \draw[process,bend left] (source.north) to node[process-label,above] {extraction\\ tool} (lod.north);
  \draw[process,bend left] (lod.north) to node[process-label,above] {query} (result.north);
  \node at (mid) [process-label,anchor=west] {evaluation\\ script};
  \draw[process] (result) to (report);
  \draw[process] (gold) to (report);
\end{scope}

      \else
      \input{SemPub_Precision_Recall_Evaluation.tikz}
      \fi
      \else
      \input{../../images/SemPub_Precision_Recall_Evaluation.tikz}
      \fi
    \end{tikzpicture}
    \captionof{figure}{Precision/recall evaluation}
    \label{fig:eval}
  \end{minipage}%
  \begin{minipage}{.40\textwidth}
    \includegraphics[width=\textwidth]{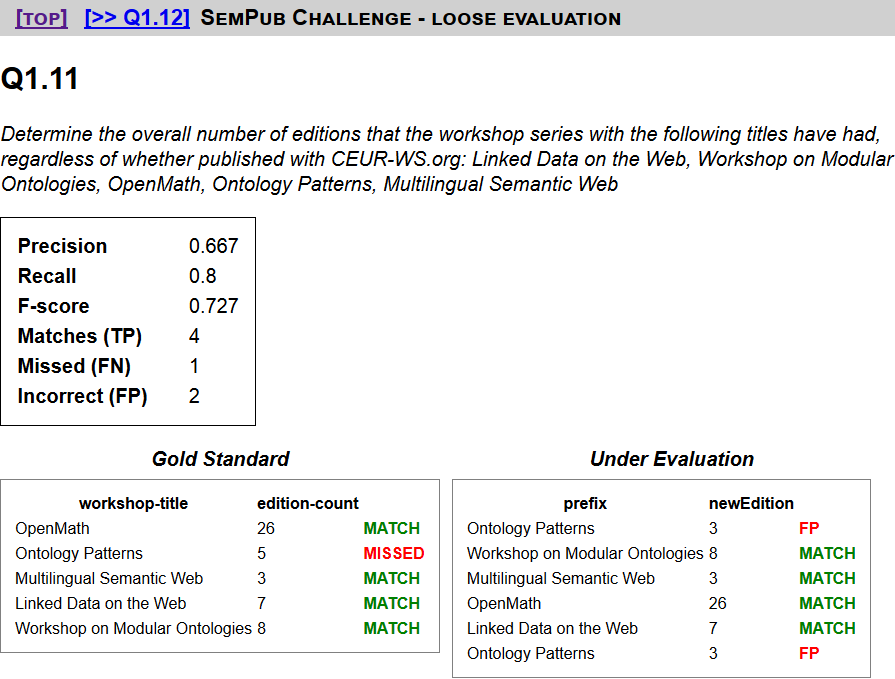}
    \captionof{figure}{Report for one query}
    \label{fig:report}
  \end{minipage}
\end{figure}

The given queries contained placeholders, e.g.\ 
“all authors of the paper titled $T$”.
For training, we specified the results expected after substituting certain values from TD for the variables.
We evaluated by substituting further values, mostly values that were only available in ED.
We defined easy as well as challenging queries, all weighted equally, to help participants get started, without sacrificing our ability to clearly distinguish the best-performing approach.
A collection of PHP scripts\footnote{\url{https://github.com/angelobo/SemPubEvaluator}} helped to automate the evaluation: they compared a CSV form of the results of the participants' SPARQL queries over their data against a gold standard of expected results
, and compiled 
a
report with 
measures and a list of false positives and false negatives (see Figures~\ref{fig:eval} and \ref{fig:report}).

\section{Task 1: Extraction and Assessment of Workshop Proceedings Information}
\label{sec:task1}

\subsection{Motivation and Objectives}
\label{sec:objectives1}

Common questions related to the quality of a scientific workshop or conference include whether a researcher should submit a paper to it or accept an invitation to its program committee, whether a publisher should publish its proceedings, or whether a company should sponsor it~\cite{BrylEtAl:SePublica2014}.
Moreover, knowing the quality of an event helps to assess the quality of the papers accepted there.
In the 2014 Challenge, we had designed Task~1 to extract from selected CEUR-WS.org workshop proceedings volumes RDF that would enable the computation of certain indicators for the workshops' quality~\cite{LangeDiIorio:SemPub14}.
The second objective of this effort was to bootstrap the publication of \emph{all} CEUR-WS.org workshops – more than 1,400 at the time of this writing – as linked data.
As discussed above in Section~\ref{sec:chall-defin-publ}, we reused the 2014 queries, with two exceptions.
As only one of the three 2014 submissions had addressed the two Task~1 queries that required metadata extraction from the PDF full text of the papers (cf. \cite{KolKoz:SemPub14}), and as Task~2 focused on full-text extraction anyway, we replaced these queries (Q1.19 and Q1.20) by similar queries that only relied on information available from HTML sources.

\subsection{Data Source}
\label{sec:data1}

The input dataset for Task~1 consists of HTML documents at different levels of encoding quality and semantics.

\begin{itemize}
\item one HTML~4 index page linking to \emph{all} workshop proceedings volumes (\url{http://ceur-ws.org/}; invalid, somewhat messy but still uniformly structured)
\item the HTML tables of contents of selected volumes.
  Their format is largely uniform but has gained more explicit structural semantics over time, while old volumes remained unchanged.
  Microformat annotations were introduced with Vol-559 in 2010 and subsequently extended, to enable automatic indexing by DBLP~\cite{DBLP}.
  RDFa (in addition to microformats) was introduced with Vol-994 in 2013, but its use is optional, and therefore it has been used in less than 10~\% of all volumes since then.
  Valid HTML5 has been mandatory since Vol-1059 in 2013; before, hardly any volume was completely valid.
\end{itemize}
Challenges in processing tables of contents include the lack of standards for marking up editors' affiliations, invited talks, and further cases described in~\cite{LangeDiIorio:SemPub14}.

The training and evaluation datasets TD1 and ED1, available at \url{https://github.com/ceurws/lod/wiki/Task1}, balance different document formats.
To enable reasonable quality assessment, TD1 already comprised certain complete workshop series, including, e.g., Linked Data on the Web, and, for some conferences, e.g., WWW 2012, all of its workshops that published with CEUR-WS.org.
In ED1, some more workshop series and conferences were completed.

\begin{table}
\centering

\footnotesize

\caption{Task 1 Data Sources}
\renewcommand{\tabularxcolumn}[1]{>{\arraybackslash}m{#1}}
\newcolumntype{Y}{>{\centering\arraybackslash}X}
\newcolumntype{Z}{>{\arraybackslash}X}

\begin{tabularx}{\textwidth}{ >{\hsize=1\hsize}Z  >{\hsize=1\hsize}Y  >{\hsize=1\hsize}Y  }
\toprule

 & Training Dataset (TD1) & Evaluation Dataset (ED1) \\
\midrule
Proceedings volumes & 98 &  148 (98 + 50)\\
… including metadata of & 1,700+ papers &  2,400+ papers \\
\midrule
Volumes using RDFa & 6 & 12 (6 + 6)\\
… using microformats only & 68 & 106 (68 + 38) \\
\bottomrule

\end{tabularx}

\label{table:Task1DataSourcesTable}
\end{table}

\subsection{Queries}
\label{sec:queries1}

The queries were roughly ordered by increasing difficulty.
Most queries from Q1.5 onward correspond to quality indicators discussed in Section~\ref{sec:objectives1}; Q1.1–Q1.4 were intended to help the participants get started.
Further background about Q1.1–Q1.18, which we reused from 2014, can be found in~\cite{LangeDiIorio:SemPub14}.
\begin{description}
\item[Q1.1] List the full names of all \textbf{editors of the proceedings of workshop} $W$.
\item[Q1.2] Count the \textbf{number of papers in workshop} $W$.
\item[Q1.3] List the full names of all \textbf{authors who have \mbox{(co-)}au\-thored a paper} in workshop $W$.
\item[Q1.4] Compute the \textbf{average length of a paper} (in pages) in workshop $W$.
\item[Q1.5 (publication turnaround)] Find out whether the proceedings of workshop $W$ were published on CEUR-WS.org before the workshop took place.
\item[Q1.6 (previous editions of a workshop)] Identify all editions that the workshop series titled $T$ has published with CEUR-WS.org.
\item[Q1.7 (chairs over the history of a workshop)] Identify the full names of those chairs of the workshop series titled $T$ that have so far been a chair in every edition of the workshop published with CEUR-WS.org.
\item[Q1.8 (all workshops of a conference)] Identify all CEUR-WS.org proceedings volumes in which 
workshops of conference $C$ in year $Y$ were published.
\item[Q1.9] Identify those papers of workshop $W$ that were \textbf{(co-)authored by at least one chair} of the workshop.
\item[Q1.10] List the full names of all \textbf{authors of invited papers} in workshop $W$.
\item[Q1.11] Determine the \textbf{number of editions} that the workshop series titled $T$ has had, regardless of whether published with CEUR-WS.org.
\item[Q1.12 (change of workshop title)] Determine the title (without year) that workshop $W$ had in its first edition.
\item[Q1.13 (workshops that have died)] Of the workshops of conference $C$ in year $Y$, identify those that did not publish with CEUR-WS.org in the following year (and that therefore probably no longer took place).
\item[Q1.14 (papers of a workshop published jointly with others)] Identify the papers of the workshop titled $T$ (which was published in a joint volume $V$ with other workshops).
\item[Q1.15 (editors of one workshop published jointly with others)] List the full names of all editors of the proceedings of the workshop titled $T$ (which was published in a joint volume $V$ with other workshops).
\item[Q1.16] Of the workshops that had editions at conference $C$ both in year $Y$ and $Y+1$, identify the \textbf{workshop(s) with the biggest percentage of growth} in their number of papers.
\item[Q1.17 (change of conference affiliation)] Return the acronyms of those workshops of conference $C$ in year $Y$ whose previous edition was co-located with a different conference series.
\item[Q1.18 (change of workshop date)] Of the workshop series titled $T$, identify those editions that took place more than two months later/earlier than the previous edition published with CEUR-WS.org.
\item[Q1.19 (institutional diversity and internationality of chairs)] Identify the affiliations and countries of all editors of the proceedings of workshop $W$.
\item[Q1.20 (continuity of authors)] Identify the full names of those authors of papers in the workshop series titled $T$ that have so far (co-)authored a paper in every edition of the workshop published with CEUR-WS.org.
\end{description}

Q1.5 (partly), Q1.12, Q1.13, Q1.16 and Q1.17 relied on the main index.

As Task~1 also aimed at producing linked data that we could eventually publish at CEUR-WS.org, the participants were additionally asked to follow a uniform URI scheme: \url{http://ceur-ws.org/Vol-NNN/} for volumes, and \url{http://ceur-ws.org/Vol-NNN/\#paperM} for a paper having the filename \url{paperM.pdf}.

\subsection{Accepted Submissions and Winners}
\label{sec:submissions1}

We received and accepted four submissions that met the requirements.

Milicka/Burget~\cite{MilBur:SemPub15}, the only new team, took advantage of the facts that, despite changes in \emph{markup}, the visual \emph{layout} of the proceedings volumes has hardly changed over 20 years, and that \emph{within} one volume non-standard layout/formatting choices are applied consistently.
They do not rely on the microformat markup at all.
The generic part of their data model covers a page's box layout and the segments of these boxes, which get tagged after text analysis.
Further domain-specific analysis yields a logical tree structure, which is finally mapped to the desired output vocabulary.
This submission won both awards: for the \emph{most innovative approach} and for the \emph{best performance}.

The three teams that had participated in 2014 evolved their submissions.
The following overview focuses on new functionality; otherwise, we refer to the 2014 overview~\cite{LangeDiIorio:SemPub14}.
Kolchin et al.~\cite{KolEtAl:SemPub15} (2014: \cite{KolKoz:SemPub14}) enriched their knowledge representation and optimised precision by adding post-processing steps including name disambiguation.
Heyvaert et al.~\cite{HeyEtAl:SemPub15} (2014: \cite{DimouEtAl:SemPub14}) simplified their HTML→RDF mapping definitions thanks to improvements of the RML mapping language, and optimised precision and recall by running systematic tests over the output to reduce failure due to, e.g., malformed literals.
Ronzano et al.~\cite{RonEtAl:SemPub15} (2014: \cite{RonzanoEtAl:SemPub14}) consulted additional external datasets and web services to support information extraction (e.g.\ the EU Open Data Portal for names of institutions) and improved their heuristics for validating, sanitising and normalising the data extracted.
Their original submission performs poorly because they forgot the trailing slash of the volume URIs.
We fixed this mistake to improve comparability.

\begin{table}
\centering

\footnotesize

\caption{Task~1 evaluation results}

\begin{tabular}{p{.24\textwidth}p{.12\textwidth}p{.11\textwidth}p{.07\textwidth}p{.1\textwidth}p{.11\textwidth}p{.11\textwidth}p{.07\textwidth}}
\toprule
Authors & \textbf{Overall\newline average\newline precision} & \textbf{Overall\newline average\newline recall} & \textbf{Ov.\newline avg.\newline F1} & Queries\newline attemp\-ted & Average\newline precision\newline on these & Average\newline recall\newline on these & Avg.\ F1 \\
\midrule
Milicka/Burget~\cite{MilBur:SemPub15}
& 0.774 & 0.591 & 0.64 & 1–20   & 0.774 & 0.591 & 0.64 \\ 
Kolchin et al.~\cite{KolEtAl:SemPub15}
& 0.658 & 0.531 & 0.565 & 1–18           & 0.731 & 0.591 & 0.628 \\ 
Heyvaert et al.~\cite{HeyEtAl:SemPub15}
& 0.254 & 0.248 & 0.244 & 1–18, 20 & 0.268 & 0.261 & 0.257 \\ 
Ronzano et al.~\cite{RonEtAl:SemPub15}
& 0.028 & 0.046 & 0.034 & 1–12, & 0.039 & 0.066 & 0.048 \\ 
… with fixed URIs
& 0.375 & 0.290 & 0.302 & 14–15 & 0.536 & 0.414 & 0.432 \\ 
\bottomrule
\end{tabular}

\label{table:Task1Eval}
\end{table}

\begin{table}
\centering

\footnotesize

\caption{Task~1 comparison to 2014 (Q1–Q18)}

\begin{tabular}{p{.25\textwidth}p{.11\textwidth}p{.1\textwidth}p{.08\textwidth}p{.07\textwidth}p{.11\textwidth}p{.1\textwidth}p{.08\textwidth}}
\toprule
2015 & Average\newline precision & Average\newline recall & Avg.\ F1 & 2014 & Average\newline precision & Average\newline recall & Avg.\ F1 \\
\midrule
Milicka/Burget~\cite{MilBur:SemPub15}
& 0.805 & 0.603 & 0.657 & n/a   &  &  \\
Kolchin et al.~\cite{KolEtAl:SemPub15}
& 0.731 & 0.591 & 0.628 & \cite{KolKoz:SemPub14} & 0.678 & 0.628 & 0.644 \\
Heyvaert et al.~\cite{HeyEtAl:SemPub15}
& 0.283 & 0.276 & 0.271 & \cite{DimouEtAl:SemPub14} & 0.153 & 0.103 & 0.117 \\
Ronzano et al.~\cite{RonEtAl:SemPub15}
& 0.031 & 0.051 & 0.037 & & \\
… with fixed URIs
& 0.417 & 0.322 & 0.336 & \cite{RonzanoEtAl:SemPub14} & 0.372 & 0.348 & 0.319 \\
\bottomrule
\end{tabular}

\label{table:Task1Comp}
\end{table}

\subsection{Lessons Learnt}
\label{sec:lessons1}

The four Task~1 submissions followed different technical approaches.
Two solutions were solely developed to address this Challenge~\cite{KolEtAl:SemPub15,RonEtAl:SemPub15}, whereas Heyvaert et al.\ and Milicka/Burget defined task-specific mappings in an otherwise generic framework~\cite{HeyEtAl:SemPub15,MilBur:SemPub15}.
The performance \emph{ranking} of the three tools evolved from 2014 has not changed (cf.\ Table~\ref{table:Task1Eval}), but their performance has improved (cf.\ Table~\ref{table:Task1Comp}) – except for Kolchin et al., who improved precision but not recall.
Disregarding the two queries that were new in 2015, the tool by Kolchin et al., which had won the best performance award in 2014, performs almost as well as Milicka's/Burget's.

In 2014, we had made first experiments with rolling out the tool by Kolchin et al.\ at CEUR-WS.org\footnote{Licensing issues slowed down progress: from Vol-1265 the metadata are open under CC0, whereas for older volumes CEUR-WS.org does not have the editors' explicit permission to republish derivatives such as extracted RDF.
Opinions diverge on the copyrightability of metadata~\cite{Coyle:MetadataCopyright2013}; DBLP actually republishes CEUR-WS.org metadata under ODC-BY.
Still, CEUR-WS.org decided not to publish old metadata under their domain; instead, \emph{we} will publish them as an outcome of this Challenge.}, but will now also evaluate Milicka's/Burget's tool.
Its reliance on the layout (which hardly ever changes) rather than the underlying markup (which improves every few years) promises low maintenance costs.

\section{Task 2: Extracting contextual information from the PDF full text of the papers}
\label{sec:task2}

\subsection{Motivation and Objectives}\label{Task2Motivation}

Task 2 was designed to test the ability to extract data from the full text of the papers. It follows last year's Task 2, which focused on extracting information about citations. 
The rationale was that the network of citations of a paper – including papers citing it or cited by that paper – is an important dimension to assess its relevance and to contextualise it within a research area.

This year we included further \emph{contextual information}. 
Scientific papers are not isolated units.  Factors that directly or indirectly contribute to the origin and development of a paper include citations, the institutions the authors are affiliated to, funding agencies, and the venue where a paper was presented. 
Participants had to make such information explicit and exploit it to answer queries providing a deeper understanding of the context in which papers were written.

The dataset's \emph{format} is another difference from 2014. 
Instead of XML sources, we used PDF this year, taken from CEUR-WS.org.
PDF is still the predominant format for publishing scientific papers, despite being designed for printing. 
The internal structure of a PDF paper does not correspond to the logical structure of its content, rather to a sequence of layouting and formatting commands. 
The challenge for participants was to recover the logical structure, to extract contextual information, and to represent it as semantic assertions.

\subsection{Data Source}\label{Task2DataSource}

The construction of the input datasets was driven by the idea of covering a wide spectrum of cases. 
The papers were selected from 21 different workshops published with CEUR-WS.org.  As these workshops had defined their own rules for submissions,
the dataset included papers in the LNCS and ACM formats.

Even if all papers 
had used the same style, their internal structures differed nevertheless. For instance, some papers used numbered citations, others used the APA 
or other styles.
Data about authors and affiliations used heterogeneous structures, too.
Furthermore, the papers used different content structures and different forms to express acknowledgements and to refer to entities in the full text (for instance, when mentioning funding, grants, projects, etc.).
 
The datasets TD2 (training) and ED2 (evaluation) are available at 
\url{https://github.com/ceurws/lod/wiki/Task2}, as a list of PDF files grouped by proceedings volume.
Table~\ref{table:Task2DataSourcesTable} reports some statistics about these datasets. TD2 is a \emph{randomly} chosen subset of papers from ED2. 
The final evaluation was performed on a randomly chosen subset of ED2 too. 
To cover all queries and balance results, we clustered input papers around each query and selected some of them from each cluster. 
Each cluster was composed of papers containing enough information to answer each query, and structuring that information in different ways.

\begin{table}[h!]
\centering

\footnotesize

\caption{Task 2 Data Sources}
\renewcommand{\tabularxcolumn}[1]{>{\arraybackslash}m{#1}}
\newcolumntype{Y}{>{\centering\arraybackslash}X}
\newcolumntype{Z}{>{\arraybackslash}X}

\begin{tabularx}{\textwidth}{ >{\hsize=.6\hsize}Y  >{\hsize=1.2\hsize}Y  >{\hsize=1.2\hsize}Y  }
\toprule

 & Training Dataset (TD2) & Evaluation Dataset (ED2) \\
 \midrule

Workshops & 12 & 21 (12 + 9)  \\

Papers & 103 (28 ACM + 75 LNCS) &  185 (103 + 22 ACM + 60 LNCS) \\
 
\bottomrule

\end{tabularx}

\label{table:Task2DataSourcesTable}
\end{table}

\subsection{Queries}\label{Task2Queries}

Our ten queries are not meant to be exhaustive but 
to cover a large spectrum of information.
The first two collect information about authors' affiliations:

\begin{description}
\item [Q2.1] Identify the \textbf{affiliations} of the authors of paper $X$
\item [Q2.2]  Identify the \textbf{papers presented at workshop $X$} and written by researchers affiliated to an \textbf{organisation located in country $Y$}
\end{description}

Affiliations can be associated to authors in different ways: listed right after the author names, placed in footnotes, or placed in a dedicated space of the paper, and so on. 
The correct identification of affiliation and authors is tricky and opens complex issues of content normalisation and homonymity management. We adopted a simplified approach: participants were required to extract all information available in the input dataset and to normalise content.

Citations are key components of the context of a paper.  Three queries deal with extracting data from bibliographies and filtering them by venue and year: 

\begin{description}
\item [Q2.3] Identify all \textbf{works cited} by paper $X$
\item [Q2.4] Identify all \textbf{works cited} by paper $X$ and published \textbf{after year $Y$}.
\item[Q2.5] Identify all \textbf{journal papers cited} by paper $X$
\end{description}

As in 2014, we some queries covered research funding. Such information is useful to investigate how funding was connected to, or even influenced, the research reported in a paper. Awareness of funding might influence the credibility and authoritativeness of a scientific work. 
The following two queries could be answered by parsing acknowledgements or other dedicated sections:

\begin{description}
\item[Q2.6] Identify the \textbf{grant(s)} that supported the research presented in paper $X$ (or part of it)
\item[Q2.7] Identify the \textbf{funding agencies} that funded the research presented in paper $X$ (or part of it)
\end{description}

Research papers often result from large projects.  Knowing them can help to better understand the scope and goal of a given work.   The following query related to projects is distinct from the previous ones as these projects are peculiar and clearly identified in the papers (usually in the acknowledgements or in footnotes): 

\begin{description}
\item[Q2.8] Identify the \textbf{EU project(s)} that supported the research presented in paper $X$ (or part of it).
\end{description}

The last two queries were meant to test entity recognition from the papers' textual content. We focused on \emph{ontologies}, as most papers in the dataset were about Semantic Web and formal reasoning and we expected ontologies to be clearly identifiable. For simplicity, we limited the search to the abstracts:

\begin{description}
\item [Q2.9] Identify \textbf{ontologies mentioned} in the abstract of paper $X$
\item [Q2.10]  Identify \textbf{ontologies introduced} in paper $X$ (according to the abstract)
\end{description}

Note that we differentiated two queries: identifying all ontologies mentioned in the abstract vs.\ those introduced for the first time in the paper (again, search was limited to the abstract). We expected participants to analyse the text and to interpret the verbs used by the authors. Nonetheless the last five queries still proved difficult and only a few were answered correctly (see below for details).

\subsection{Accepted Submissions and Winner}\label{Task2Submissions}

We received six submissions for Task 2:

Sateli/Witte~\cite{SatWit:SemPub15} proposed a rule-based approach.
They composed two logical components in a pipeline: a syntactic processor 
to identify the basic layout units and to cluster them into logical units, and a semantic processor 
to identify entities in text by pattern search.
The framework is based on GATE and includes an RDF mapper that 
transforms the extracted data into RDF triples. 
The mapper's high flexibility contributed to this submission winning the \emph{most innovative approach award}.

Tkaczyk/Bolikowski~\cite{TkaBol:SemPub15} won the \emph{best performing tool award} for their CERMINE framework: a Java application extracting metadata from scientific papers by supervised and unsupervised machine learning.
The tool was successfully used for the Challenge with a few modifications, including the implementation of an RDF export. 
It performed extremely well in extracting affiliations and citations.

Klampfl/Kern~\cite{KlaKer:SemPub15} also used supervised and unsupervised machine learning. 
Their modular framework identifies and clusters building blocks of the PDF layout.
Trained classifiers helped to detect the role of each block (authorship data, affiliations, etc.).
The authors built an ontology of computer science concepts and exploited it for the automatic annotation of funding, grant and project data.

Ronzano et al.~\cite{RonEtAl:SemPub15} extended their Task~1 framework to extract data from PDF. 
Their pipeline includes text processing and entity recognition modules and employs external services for mining PDF articles, and to increase the precision of the citation, author and affiliation extraction. 

Integrating multiple techniques and services is also a key aspect of MACJa, the system presented by Nuzzolese et al.~\cite{NuzEtAl:SemPub15}.
Mainly written in Python and Java, it extracts the textual content of PDF papers using PDFMiner and runs multiple analyses on that content. 
Named Entity Recognition (NER) techniques help to identify authors and affiliations; CrossRef APIs are queried to extract data from citations; 
NLP techniques, pattern matching and alignment to lexical resources are finally exploited for detecting ontologies, grants and funding agencies.

Kovriguina et al.~\cite{KovEtAl:SemPub15} presented a simple but efficient architecture, implemented in Python and sharing code with their Task~1 submission~\cite{KolEtAl:SemPub15}. 
Their approach is mainly based on templates and regular expressions and relies on some external services for improving the quality of the results (e.g., DBLP for checking authors and citations).  An external module extracts the plain text from PDFs. This text is matched against a set of regular expressions to extract the relevant parts; the serialisation in RDF follows a custom ontology derived from BIBO.

Table~\ref{table:Task2Eval} summarises the results of the performance evaluation.

\begin{table}
\centering

\footnotesize

\caption{Task~2 evaluation results}

\begin{tabular}{p{.30\textwidth}p{.14\textwidth}p{.13\textwidth}p{.13\textwidth}}
\toprule
Authors & \textbf{Precision} & \textbf{Recall} & \textbf{F1 score}\\
\midrule
Tkaczyk/Bolikowski~\cite{TkaBol:SemPub15}
 & 0.369 & 0.417 & 0.381 \\ 

Klampfl/Kern~\cite{KlaKer:SemPub15}
& 0.388 & 0.285 & 0.292 \\ 

Nuzzolese et al.~\cite{NuzEtAl:SemPub15}
& 0.274 & 0.251& 0.257 \\ 

Sateli/Witte~\cite{SatWit:SemPub15}
& 0.3 & 0.252 & 0.247 \\ 

Kovriguina et al.~\cite{KovEtAl:SemPub15}
& 0.289 & 0.3 & 0.265 \\ 

Ronzano et al.~\cite{RonEtAl:SemPub15}
& 0.316 & 0.401 & 0.332 \\ 

\bottomrule

\end{tabular}

\label{table:Task2Eval}
\end{table}

\subsection{Lessons Learnt}\label{Task2Lessons}

We see two main reasons for the unexpectedly low performance:

\textbf{The complexity of the task.}
When designing the task, we decided to explore a larger amount of contextual information to identify the most interesting issues in this area.
In retrospect, this choice led us to defining a difficult task, which instead could have been structured differently. 
The queries are logically divided in two groups: queries Q2.1--Q2.5 required participants to identify logical units in PDFs; the others required additional content processing. 
As these two blocks required different skills, we could have separated them in two tasks.
Queries within one group, however, were perceived as too heterogeneous.
For next year, we are considering fewer types of queries with more cases each.

\textbf{The evaluation.}  As we considered only some papers for the final evaluation (randomly selected among those in the evaluation dataset) some participants were penalised: their tool could have worked well on other values, which were not taken into account. \todo{CL@AdI: please give one concrete example}Some low scores also depended on imperfections in the output format. Since the evaluation was fully automated -- though the content under evaluation was normalised and minor differences were not considered errors -- these imperfections impacted results negatively.

\section{Task 3: Interlinking}
\label{sec:task3}

\subsection{Motivation and Objectives}
\label{Task3Motivation}

Task~3 was newly designed to assess the ability to identify same entities across different datasets of the same domain, thus establishing links between these datasets.
Participants had to make such links explicit 
and exploit them to answer comprehensive queries about events and persons. 
The CEUR-WS.org data in itself provide incomplete information about conferences and persons.
This information can be complemented by interlinking the dataset with others to broaden the context and to allow for more reliable conclusions about the quality of scientific events and the qualification of researchers.


\subsection{Data Source}
\label{Task3DataSource}

The input for Task~3 consists of datasets 
in different RDF serialisations and different levels of encoding quality and semantics.
For each dataset, we made an RDF dump and an endpoint or Triple Pattern Fragments~\cite{VerborghEtAl:TPF14} available.
The complete training dataset TD3, available at 
\url{https://github.com/ceurws/lod/wiki/Task3}, 
comprises multiple individual datasets accessible in different ways:

\begin{description}
\item[CEUR-WS.org\label{CEURWS}]
This dataset includes the workshop proceedings volumes up to Vol-1322; it was produced in January 2015 by Maxim Kolchin using his extraction tool, which had won Task~1 of the 2014 Challenge~\cite{KolKoz:SemPub14}.
\begin{description}
\item[RDF dump]~{\footnotesize\url{https://github.com/ceurws/lod/blob/master/data/ceur-ws.ttl}}
\item[Triple Pattern Fragments]{\footnotesize\url{http://data.linkeddatafragments.org/ceur-ws}}
\end{description}

\item[COLINDA\label{COLINDA}]
The Conference Linked Data\footnote{\url{http://www.colinda.org/}} 
dataset  exposes metadata about scientific events (conferences
and workshops) announced at 
EventSeer and WikiCfP\footnote{See \url{http://eventseer.net/} and \url{http://www.wikicfp.com/}} from 2002.
COLINDA includes information about the title, description, date and venue of events.
It is interlinked with DBLP (see below), Semantic Web Dog Food (see below), 
GeoNames and DBpedia\footnote{See \url{http://www.geonames.org/} and \url{http://dbpedia.org/}}.
\begin{description}
\item[RDF dump]~{\footnotesize\url{https://github.com/ceurws/lod/blob/master/data/colinda.nt}}
\item[Endpoint]~{\footnotesize\url{http://data.colinda.org/endpoint.html}}
\item[Triple Pattern Fragments]{\footnotesize\url{http://data.linkeddatafragments.org/colinda}}
\end{description}

\item[DBLP\label{DBLP}] 
The DBLP computer science bibliography~\cite{DBLP} is the prime reference 
for open bibliographic information on computer science publications.
It currently indexes over 2.6 million publications by more than 1.4 million authors, 
in more than 25,000 journal volumes, 24,000 conferences or workshops, and 17,000 monographs. 
We used the DBLP++ dataset\footnote{\url{http://dblp.l3s.de/dblp++.php}}.
\begin{description}
\item[RDF dump]~{\footnotesize\url{http://dblp.l3s.de/dblp-2015-02-14.sql.gz}}
\item[Endpoint]~{\footnotesize\url{http://dblp.l3s.de/d2r/sparql}}
\item[Triple Pattern Fragments]~{\footnotesize\url{http://data.linkeddatafragments.org/dblp}}
\end{description}

\item[Lancet\label{SLT}]
The Semantic Lancet Triplestore dataset\footnote{\url{http://www.semanticlancet.eu/}} contains 
metadata about papers published in the Journal of Web Semantics by Elsevier.
For each paper, the dataset reports bibliographic metadata, abstract and citations.
\begin{description}
\item[RDF dump]~{\footnotesize\url{https://github.com/ceurws/lod/blob/master/data/lancet.ttl}}
\item[Endpoint]~{\footnotesize\url{http://data.linkeddatafragments.org/lancet}}
\item[Triple Pattern Fragments]~{\footnotesize\url{http://data.linkeddatafragments.org/lancet}}
\end{description}

\item[SWDF\label{SWDF}]
The Semantic Web Dog Food\footnote{\url{http://data.semanticweb.org/}}
metadata covers around 5,000 papers, 11,000 people, 3,200 organisations,
45 conferences and 230 workshops.
\begin{description}
\item[RDF dump]~{\footnotesize\url{http://data.semanticweb.org/dumps/}}
\item[Triple Pattern Fragments]~{\footnotesize\url{http://data.linkeddatafragments.org/dogfood}}
\end{description}

\item[Springer LD\label{SLD}]
This dataset\footnote{\url{http://lod.springer.com/}} contains metadata of around
1,200 conference series and 8,000 proceedings volumes
published by Springer in the Lecture Notes in Computer Science (LNCS),
Lecture Notes in Business Information Processing (LNBIP),
Communications in Computer and Information Science (CCIS),
Advances in Information and Communication Technology (IFIP-AICT), and
Lecture Notes of the Institute for Computer Sciences, Social Informatics and Telecommunications Engineering (LNICST) series.
\begin{description}
\item[RDF dump]~{\footnotesize\url{https://github.com/ceurws/lod/blob/master/data/springer.nt}}
\item[Endpoint]~{\footnotesize\url{http://lod.springer.com/sparql}}
\item[Triple Pattern Fragments]~{\footnotesize\url{http://data.linkeddatafragments.org/springer}}
\end{description}

\end{description}

\subsection{Queries}
\label{Task3Queries}

The list of queries follows, ordered by increasing difficulty:

\begin{description}
\item [Q3.1 (Same entities within the CEUR-WS.org dataset)]
Identify and interlink same entities that appear with different URIs within the CEUR-WS.org dataset.
Same persons (authors and/or editors) or same events (conferences) might have been assigned different URIs.
\item [Q3.2 (Same entities across different datasets)]
Identify all different instances of the same entity in different datasets.
Same entities (persons, articles, proceedings, events) might appear in different datasets with different URIs.

\item [Q3.3 (Workshop call for papers)]
Link a CEUR-WS.org workshop $W$ to its call for papers announced on EventSeer and/or WikiCfP.

\item [Q3.4 (Workshop website)]
Link a workshop or conference $X$ that appears in the CEUR-WS.org dataset to the workshop's or conference's website URL.

\item [Q3.5 (Overall contribution to the conference)]
Identify all papers edited by an author $A$ of a CEUR-WS.org paper $P$ presented at workshop $W$ co-located with conference $C$, and who was also author of a main track paper at the same conference $C$.

\item [Q3.6 (Overall activity in a year)]
Identify, for an author $A$ of a CEUR-WS.org paper $P$, all his/her activity in year $Y$.

\item [Q3.7 (Full series of workshops)]
Identify the full series of workshop $W$ regardless of whether individual editions published with CEUR-WS.org.

\item [Q3.8 (Other co-authors)]
Identify people who co-authored with author $A$ of a paper $P$ published by CEUR-WS.org but did not co-author any CEUR-WS.org papers published in year $Y$ with him/her.

\end{description}

\subsection{Lessons Learnt}
\label{Task3Lessons}

For Task 3, we did not receive any submissions, even though participants of the other two tasks had expressed interest.
For the next challenge we are considering interlinking tasks that focus on directly extending the information extraction tasks.
This way, we expect to lower the entrance barrier for participants of the information extraction tasks to also address interlinking.

\section{Overall Lessons Learnt for Future Challenges}
\label{sec:over-less-learn}

As a result of the 2014 Semantic Publishing Challenge, we had obtained an RDF dataset about the CEUR-WS.org workshops.
This dataset served as the foundation to build the 2015 Challenge on.
We designed all three tasks around the same dataset.
This was a good choice in our opinion, as participants could extend their existing tools to perform multiple tasks, and it also opens new perspectives for future collaboration: participants' work could be extended and integrated in a shared effort for producing LOD useful for the whole community.

On the other hand, the evaluation process presented some weaknesses. One participant, for instance, suggested to use an evaluation dataset disjoint from the training dataset to avoid over-training; we should also consider a larger set of instance queries and provide users with intermediate feedback so that they can progressively refine their tools towards providing more precise results. 

The definition of the tasks presented some issues this year as well. It was difficult, in particular, to define tasks that were appealing and with balanced difficulty. In retrospect, some tasks were probably too wide and difficult.

Next year, we plan to further increase the reusability of the extracted data, e.g., by asking for an explicit representation of licensing information, but primarily we want to put more emphasis on interlinking.
For example, by linking publications to related social websites as SlideShare or Twitter, we will be able to more appropriately assess the impact of a scientific event within the community.

\paragraph{Acknowledgements}
We thank our reviewers, our sponsors Springer and Mendeley, and our participants for their hard work, creative solutions and useful suggestions.
This work has been partially funded by the European Commission under grant agreement no.\ 643410.

\printbibliography
\end{document}

